\documentclass[epj]{svjour}
\usepackage{graphics}
\usepackage{epstopdf}
\usepackage{xcolor} 
\usepackage{cite}  
\begin{document}
\title{The Great Chinese Fireball of December 22, 2020}
\author{
Albino Carbognani\inst{1}
}                     
\offprints{albino.carbognani@inaf.it}          
\institute{INAF-Osservatorio di Astrofisica e Scienza dello Spazio, Via Piero Gobetti, 93/3, 40129 Bologna, Italy
}
\date{Received: date / Revised version: date}
%
\abstract{
On December 22, 2020 at about 23h 23m 33s UTC a very bright fireball was seen in the sky of south-eastern China. The fireball lasted for several seconds and ended with an atmospheric explosion that was detected by US surveillance satellites. According to CNEOS's data, the fireball moved with a mean speed of 13.6 km/s and exploded at an altitude of about 35.5 km (Lat. $31.9^{\circ}$ N; Long. $96.2^{\circ} $ E). In this paper we estimate the atmospheric trajectory, the heliocentric orbit and the strewn fields for different mass/section ratio of the fragments. The trajectory was about from north to south and with low inclination ($5^{\circ} \pm 2^{\circ}$) with respect to the local surface. The explosion height appear consistent with a fragmented rocky body and the heliocentric orbit supports an asteroidal origin. The probable strewn field extend between two points with coordinates ($+31.3^\circ$ N; $96.3^\circ$ E) and ($+30.3^\circ$ N; $96.5^\circ$ E), for a total area of about 4000 $\textrm{km}^2$. This large extension is a direct consequence of the low inclination of the trajectory. Given the unknown uncertainty of the CNEOS's data, these results should be taken with caution.
\PACS{
      {PACS-key}{discribing text of that key}   \and
      {PACS-key}{discribing text of that key}
     }  
} 
\authorrunning{Carbognani}
\titlerunning{Great Chinese Fireball}
\maketitle

\section{Introduction}
\label{sec:intro} 
The Earth's atmosphere is constantly bombarded by macroscopic bodies belonging to the Solar System, in the meter or few meters class in diameter\cite{Devillepoix2019}. According to data made public by CNEOS (NASA-JPL Center for NEOs Studies)\footnote{https://cneos.jpl.nasa.gov/fireballs/ accessed 2020 December 29}, there were 851 bright fireballs reported by US government sensors, from April 15, 1988 to December 28, 2020. However the fireballs with also the impact speed vector - i.e. the good cases - are about 230 only. On average there is an event every two weeks anywhere on Earth. The satellites record the flare of the atmospheric explosions that occur when the body, stressed by the pressure and heat that develops during the fall, disintegrate itself. These events generally occurs at hypersonic speed, that is with a Mach number greater than about 5. The most intense atmospheric explosion ever detected was the Chelyabinsk event on February 15, 2013 \cite{Popova2013}.\\ 

On December 22, 2020 at 23:23:33 UT, a very bright fireball - which we will also call with the code ``Country Code YYYYMMDD'', i.e. CN20201222 - was seen from south China. At some point of its trajectory the fireball exploded even if, most likely, the major fragments continued the fall towards the ground. It was a Chelyabinsk-like event, albeit - fortunately - on a much smaller scale. In what follows, we want to reconstruct an identikit of the Chinese fireball based on the available CNEOS's data. About orbit, dark flight and strewn field computation we have used the same process applied to two Italian fireballs, IT20170530 \cite{Carbognani2020} and IT20200101, which led to recovery of the Cavezzo meteorite by the Italian PRISMA Project\cite{Gardiol2020}.

\begin{figure}
\begin{center}
\resizebox{0.8\textwidth}{!}{\includegraphics{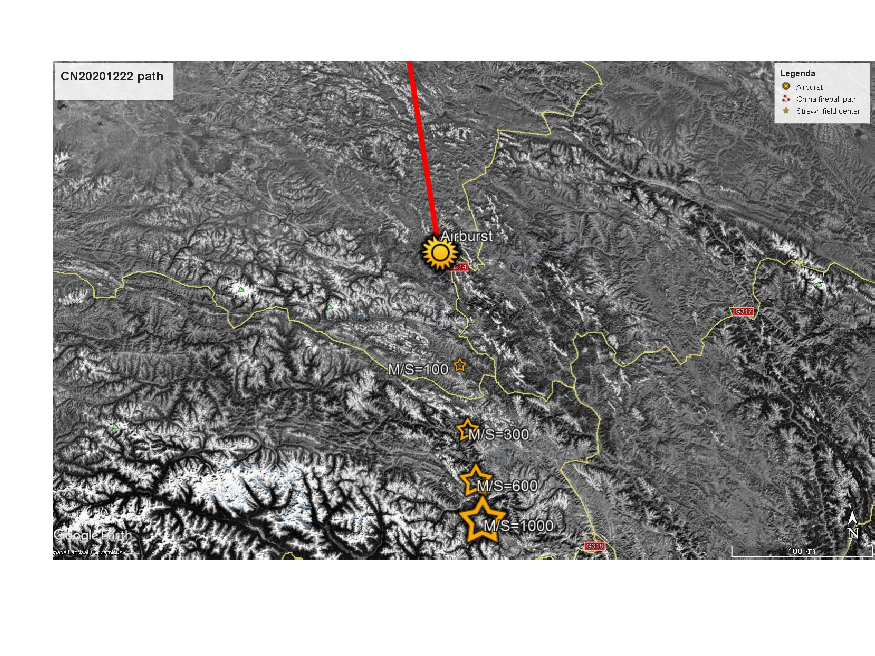}}
\end{center}
\caption{A Google Earth map showing trajectory (red line), explosion and strewn fields nominal center of CN20201222. The trajectory is determined in Section~\ref{sec:trajectory}, while the coordinates of the strewn fields center (orange stars), based on selected mass/section ratio of the hypothetical fragments, are reported in Table~\ref{tab:strewn_field_center}. The size of the stars is approximately proportional to the size of the fragment. The north is on top, the west on the left.}
\label{fig:fireball_trajectory}
\end{figure}

\section{Data collection}
\label{sec:data} 
We use satellites data from CNEOS that give fundamental information about main explosion location, height and pre-atmospheric speed vector (see Table~\ref{tab:cneos_table}). The IMO\footnote{International Meteor Organization, www.imo.net} website did not collect eyewitness accounts for this event and videos posted on YouTube are not very useful because, generally, they do not have data about the location from which they were recorded\footnote{https://www.youtube.com/watch?v=N-npJtRt2TI}. With a more in-depth investigation it will probably be possible to trace the locations. \\
In general, CNEOS's events agree in absolute time with independent records to within a few seconds and also locations are correct in most cases\cite{Devillepoix2019}. The issue of speed accuracy is more delicate. It has been found that in some cases the error on the speed was of the order of $30\%$, while the radians have an accuracy of a couple of degrees. In general, only about $40\%$ of the CNEOS events have an acceptable speed accuracy able to compute the orbit\cite{Devillepoix2019}. So, especially the results about the heliocentric orbit, should be taken with caution.\\

As a test of the uncertainties about the data provided by CNEOS, we applied the same mathematical procedure which will be used later in Section~\ref{sec:trajectory} and Section~\ref{sec:orbit_progenitor} for CN20201222 to well-known fireballs cases, i.e. whose trajectory and orbit are known independently from CNEOS's data: the Vi\~nales fireball of April 1, 2019\cite{Zuluaga2019}, the Chelyabinsk event\cite{Popova2013} and the 2008 TC3 impact\cite{Farnocchia2017}. The total Impact energy of these events are 1.4, 440 and 1 kt. The results about trajectory and orbit computation of these fireballs are shown in Table~\ref{tab:orbit_test}, together with the real values: as we can see the agreement looks good. Considering the difference between our computations - based over CNEOS's data - and the measured values in the reference papers, the mean uncertainties about azimuth, inclination and impact velocity are of the order of $\pm 2^{\circ}$, $\pm 3^{\circ}$ and $\pm 5\%$ respectively. These values appear in good agreement with the uncertainties values mentioned by Devillepoix et al.\cite{Devillepoix2019}, also much better in the case of impact speed. Obviously three cases on about 230 good fireballs are not very statistically significant but we are confident that, at least for the most energetic events, CNEOS's data are a good starting point for an event analysis.

\begin{table}
\centering
\caption{A comparison between derived from CNEOS's data and true value for best fireballs case. The analysis of these well known cases is also a general check for our computation method.}
\label{tab:orbit_test}
\begin{tabular}{lcc|cc|cc}
\hline
Quantity & CNEOS & Vi\~nales\cite{Zuluaga2019} & CNEOS & Chelyabinsk\cite{Popova2013} & CNEOS & 2008 TC3\cite{Farnocchia2017}\\
\hline
Arrival azimuth ($^\circ$)                 &   35.16   &  31.80  &   99.89   &  103.50  &  281.35  &  281.10 \\
Traj. inclination ($^\circ$)               &   176.06  &  178.90 &   15.92   &  18.55   &   18.21  &   20.83 \\
Impact speed (km/s)                        &   16.3    &  16.90  &   18.6    &  19.03   &   13.3   &   12.38 \\
Semi major axis (AU)                       &   1.20    &  1.22   &   1.71    &   1.72   &    1.33  &    1.31 \\
Eccentricity	                           &   0.38    &  0.39   &   0.56    &   0.57   &    0.34  &    0.31 \\
Orbit inclination ($^\circ$)               &   9.62    &  11.47  &   4.08    &   4.98   &    2.82  &    2.54 \\
Longitude of the ascending node ($^\circ$) &   132.33  &  132.28 &   326.42  &   326.46 &  194.09  &  194.10 \\
Longitude of perihelion ($^\circ$)         &   49.53   &  49.25  &   76.05   &   74.13  &   71.90  &   68.54 \\
Argument of Perihelion ($^\circ$)          &   277.20  & 276.97  &   109.62  &   107.67 &  237.81  &  234.44 \\
\hline
\end{tabular}     
\end{table}

Table~\ref{tab:cneos_table} summarizes all the data underlying our analysis. We assumed that the uncertainty was about $\pm 5$ s on event time, $\pm 0.1^{\circ}$ on the explosion coordinates, 0.5 km on the explosion height and about $\pm 10\%$ on the speed components. With these values of the initial uncertainties, the uncertainties on the trajectory appear very similar to those obtained for Vi\~nales, Chelyabinsk and 2008 TC3 (see Section~\ref{sec:trajectory}). On the other hand, the orbital elements uncertainties appear a little bigger because we have kept higher the speed uncertainty (see Table~\ref{tab:orbit_table}).

\begin{table*}
\centering
\caption{CN20201222 satellites data from CNEOS. In the speed vector the dominant component is along the negative z axis and this indicates a predominant movement from north to south. 
Uncertainties are assumed, see Section~\ref{sec:data}, and not provided by CNEOS.}
\label{tab:cneos_table}

\begin{tabular}{lc}
\hline
Quantity                     & Values\\
\hline
Date (YYYY-MM-DD)            &  2020-12-22 \\
UT (hh:mm:ss)                &  $23:23:33 \pm 5 s$ \\
Lat.  ($^\circ$)             &  $31.9^{\circ} \pm 0.1^{\circ}$ N   \\
Long. ($^\circ$)             &  $96.2^{\circ} \pm 0.1^{\circ}$ E   \\
Explosion height (km)        &  $35.5 \pm 0.5$          \\
Vector speed (km/s)          &  $\vec v=(-2.6\pm 0.3 ~~5.9\pm 0.6 ~~-12.1\pm 1)$    \\
Scalar speed (km/s)          &  $13.6 \pm 1 $\\
Total Impact Energy (kt)     &  $9.5 \pm 0.1$    \\
\hline
\end{tabular}
\end{table*}

\section{Atmospheric trajectory}
\label{sec:trajectory} 
Data from CNEOS give the values shown in Table~\ref{tab:cneos_table} for the main fireball explosion. The velocity $\vec v$ is the pre-impact velocity in a geocentric Earth-fixed reference frame defined as follows: the z-axis is directed along the Earth's rotation axis towards the celestial north pole, the x-axis lies in the Earth's equatorial plane, directed towards the prime meridian, and the y-axis completes the right-handed coordinate system. \\
In order to trace the atmospheric trajectory, with inclination angle and azimuth, we perform two rotation on $\vec v$. The first rotation was made clockwise in the $xy$ plane by an angle equal to the longitude $\lambda_e$ of the explosion. The second also clockwise in the $xz$ plane of an angle equal to colatitude $\theta_e=90^{\circ}-\phi_e$: 

\begin{equation}
\vec v_e=Ry(-\theta_e)\left [Rz(-\lambda_e)\vec v\right ]
\label{eq:rotation}
\end{equation}

In Eq.~\ref{eq:rotation} $Rz$ and $Ry$ are the standard rotation matrix in a plane around the z-axis and the y-axis, respectively. 
In this way we have the velocity vector seen from the explosion point, and we can easily compute the trajectory inclination respect to the Earth's surface and the azimuth where the fireball came from (counted from north toward east). To estimate the uncertainty about trajectory angles, i.e. inclination and azimuth, we performed a Monte Carlo simulation using 5000 different speed scenarios compatible with the adopted uncertainties given in Table~\ref{tab:cneos_table}.\\

The results show that the fireball followed an atmospheric trajectory inclined by an angle of $5^{\circ} \pm 2^{\circ}$ with respect to the horizontal plane, with an azimuth of
$352^{\circ} \pm 1^{\circ}$ (Fig.~\ref{fig:fireball_trajectory}). The topocentric equatorial coordinates of the apparent radiant are near the star $\gamma$ Cas in $\alpha=17^{\circ} \pm 3^{\circ}$ and $\delta=62^{\circ} \pm 3^{\circ}$ (to date), while the geocentric equatorial coordinates of the true radiant are $\alpha=10^{\circ} \pm 2^{\circ}$ and $\delta=35^{\circ} \pm 7^{\circ}$ (J2000). \\
The fireball traveled about from north to south and slowly entered the densest layers of the terrestrial atmosphere. The meteoroid could not withstand the pressure and intense heat that developed during the fall, and exploded at about 35.5 km above sea level. After the main explosion, the major fragments probably continued their fall towards the ground in dark flight phase.\\
Assuming that the kinetic energy value is equal to the value of the estimated total impact energy $T_{E}$, the equivalent meteoroid diameter $D$ is given by:

\begin{equation}
D=2\sqrt[3]{\frac{3 T_{E}}{2\pi \rho v^2}}
\label{eq:diameter}
\end{equation}

From Eq.~\ref{eq:diameter}, with $T_{E} \approx 9.5 ~\textrm{kt}\approx 4\cdot 10^{13} ~\textrm{J}$, $v\approx 13600 ~\textrm{m}/\textrm{s}$ and a bulk density $\rho\approx 2700~\textrm{kg}/\textrm{m}^3$ (typical of an S-type asteroid \cite{carbognani2017}), it follows that the Chinese fireball was generated by a small asteroid with an equivalent diameter of about 7 meters.\\
Usually the meteoroids fragmentation model assume that the fragmentation process starts when the aerodynamic pressure is equal to the mechanical strength $S$. According to the meteoroid height $h_e \approx 35.5$ km and speed $v_e\approx 13.6$ km/s in the main explosion, we can estimate a strength of about \cite{Foschini1999, Farinella2001}: 

\begin{equation}
S=\frac{\left( \gamma -1 \right)\left( 1+\alpha \right)}{2\gamma}\rho_{sl}{v_e}^2 \exp\left(-h_e/H\right)\approx 2\cdot 10^6 ~\textrm{Pascal} 
\label{eq:strength}
\end{equation}

In Eq.~\ref{eq:strength}, $\gamma\approx 3$ is the ratio of specific heats and $\alpha\approx 1$ is the coefficient of ionization for plasma state; $H\approx 8 ~\textrm{km}$ is the atmospheric scale height and $\rho_{sl}\approx 1.22 ~\textrm{kg}/\textrm{m}^3$ is the atmospheric density at sea level. Considering that for a chondrite a strength $S\approx 10^7$ Pascal could be expected\cite{Foschini1999, Devillepoix2019}, the estimated value tell us that, probably, the small asteroid of the Chinese fireball was a rocky body with some fractures or voids that decreased the strength \cite{Popova2011}.\\

\begin{figure}[h]
\begin{center}
\resizebox{0.4\textwidth}{!}{\includegraphics{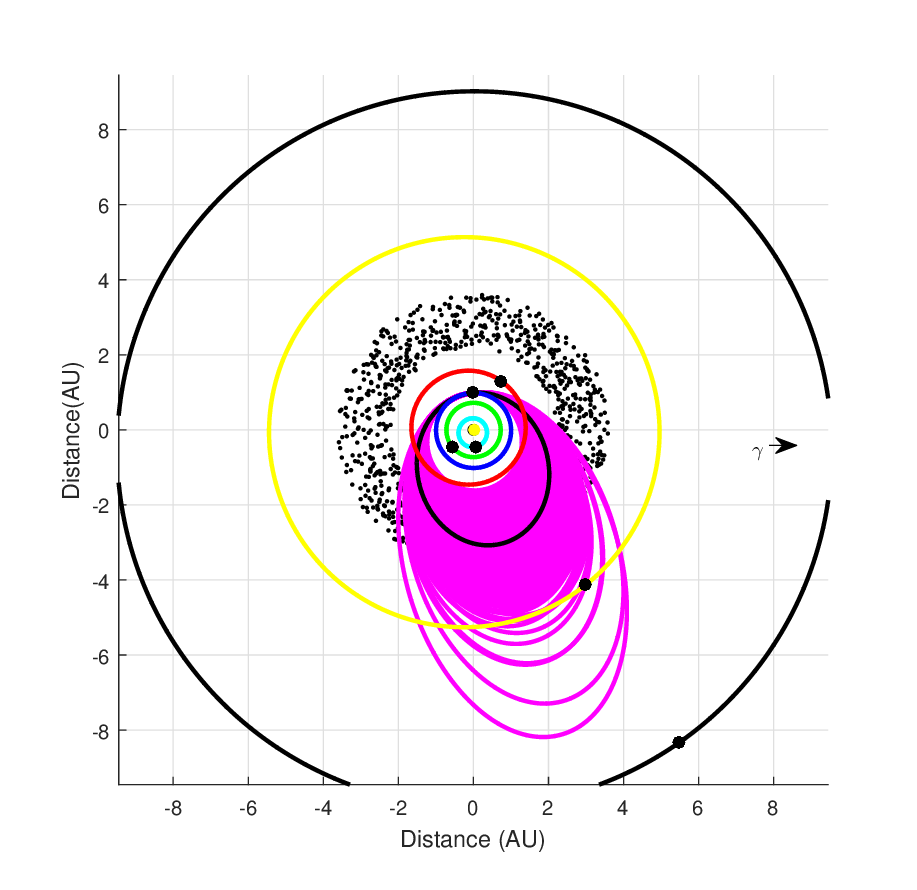}}
\end{center}
\caption{The heliocentric nominal orbit of CN20201222 projected on the Ecliptic plane (black). The orbits in magenta color are the Monte Carlo clones, useful to show the uncertainty of the nominal orbit.
Most of the possible orbits are from the Main Belt.}
\label{fig:orbit}
\end{figure}

\section{Orbit and search for a progenitor body}
\label{sec:orbit_progenitor}

In order to compute the heliocentric orbit we take the vectorial speed from Table~\ref{tab:cneos_table} as a pre-atmospheric value and correcting it for the Earth's rotation and gravity using ``zenith attraction'' method\cite{Ceplecha1987, Carbognani2020}. Taking into account the true distance and position of the Earth relative to the Sun, we found the heliocentric orbit whose elements are shown in Table~\ref{tab:orbit_table}. In order to estimate the uncertainty of the orbital elements, a Monte Carlo approach with 500 clones was performed. The computed orbital elements indicate that the meteoroid was an Apollo-type object, with aphelion in the middle of the Main Belt and with low inclination above the Ecliptic plane. The nominal orbit has a Tisserand invariant respect to Jupiter equal to $T_J = 3.6 \pm 0.5$, a typical asteroid value. Doing the statistic of the Tisserand invariants for all the clones it results that - with a probability of about 90\%  - the progenitor was moving on an asteroidal orbit (see Fig.~\ref{fig:orbit}). \\
In order to identify a possible parent body among the known NEAs, we use the $D_N$ criterion introduced by Valsecchi et al.\cite{Valsecchi1999} for meteoroid streams identification which is based on meteoroid's encounter conditions with Earth:

\begin{equation}
    D_N = \sqrt{{\left( U_{nea}-U_{cn} \right)}^2+{\left( \cos\theta_{nea}-\cos\theta_{cn} \right)}^2+{\left( 2\sin\frac{\phi_{nea}-\phi_{cn}}{2}\right)}^2+{\left( 2\sin\frac{\lambda_{nea}-\lambda_{cn}}{2}\right)}^2}
    \label{eq:valsecchi}
\end{equation}

In Eq.~\ref{eq:valsecchi} the index $cn$ refers to the fireball, while $nea$ to the NEA. Also $U$ represents the normalized object velocity in a geocentric reference frame; $\theta$ and $\phi$ are two angles defined by the geocentric velocity vector, related to the radiant in the case of a meteor stream while $\lambda$ is the longitude of the crossing point along the Earth orbit, that is, the equivalent to the date of the crossing. This criterion uses geocentric quantities and two of the quantities that are used in $D_N$ (i.e. $U$ and $cos\theta$), have been shown to be nearly invariant under the secular perturbation\cite{Gronchi2001}. Many factors influence the dynamical evolution of a meteoroid, and some of them result from forces other than gravitation, especially for meteoroids of very small size. However, over not too long time-scales, and in the absence of planetary close encounters, we can assume that only planetary secular perturbations affect meteoroid orbits. For this reason we consider the $D_N$ criterion useful for progenitor search. We refer the reader to Valsecchi et al.
\cite{Valsecchi1999} for more details. \\
For the NEAs with a not too short orbital arc, the relevant quantities of Eq.~\ref{eq:valsecchi} are conveniently tabulated by NEODyS\footnote{https://newton.spacedys.com/~neodys2/propneo/encounter.cond}, while for CN20201222 nominal orbit we have 
$U_{cn} = 0.26$, $\theta_{cn} = 34.9^{\circ}$, $\phi_{cn} = 215.9^{\circ}$ and $\lambda_{cn}=91.3^{\circ}$ (the longitude of the Earth at the time of fall). We looked for NEAs for which $D_N \leq 0.15$ and we found the following objects as a candidate progenitor: 2006 BP7	($D_N = 0.14$), 2010 CM1 ($D_N = 0.14)$, 2020 KC6 ($D_N = 0.14)$ and 2021 CR4 with ($D_N = 0.15$). None of these NEAs have a particularly short distance $D_N$ from the Chinese fireball, indeed they are all about at the same distance. We conclude that the parent NEA of CN20201222 probably is not between the known NEAs.

\begin{table}
\centering
\caption{Data about the asteroid heliocentric orbit. The standard deviations are obtained with a Monte Carlo computation with 500 clones. The longitude of the ascending node has very low uncertainty because the value is determined by the time of the fireball fall only.}
\label{tab:orbit_table}
\begin{tabular}{lc}
\hline
Quantity & Numerical value (J2000)\\
\hline
Semi major axis (AU)                       & $2.0 \pm 0.4$ \\
Eccentricity	                           & $0.5 \pm 0.1$\\
Orbital Period (years)	                   & $2.9 \pm 0.9$\\
Orbit inclination ($^\circ$)               & $6 \pm 2$\\
Longitude of the ascending node ($^\circ$) & $271.30123 \pm 0.00006$\\
Longitude of perihelion ($^\circ$)         & $104 \pm 10$\\
Argument of Perihelion ($^\circ$)          & $192 \pm 9$\\
Perihelion passage (JD)	                   & $2458145 \pm 355$\\
Perihelion distance (AU)                   & $0.976 \pm 0.004$\\
Aphelion distance (AU)                     & $3.1 \pm 0.8$\\
\hline
\end{tabular}     
\end{table}

\section{Dark flight and strewn field}
\label{sec:field} 
\begin{figure}[h]
\begin{center}
\resizebox{0.4\textwidth}{!}{\includegraphics{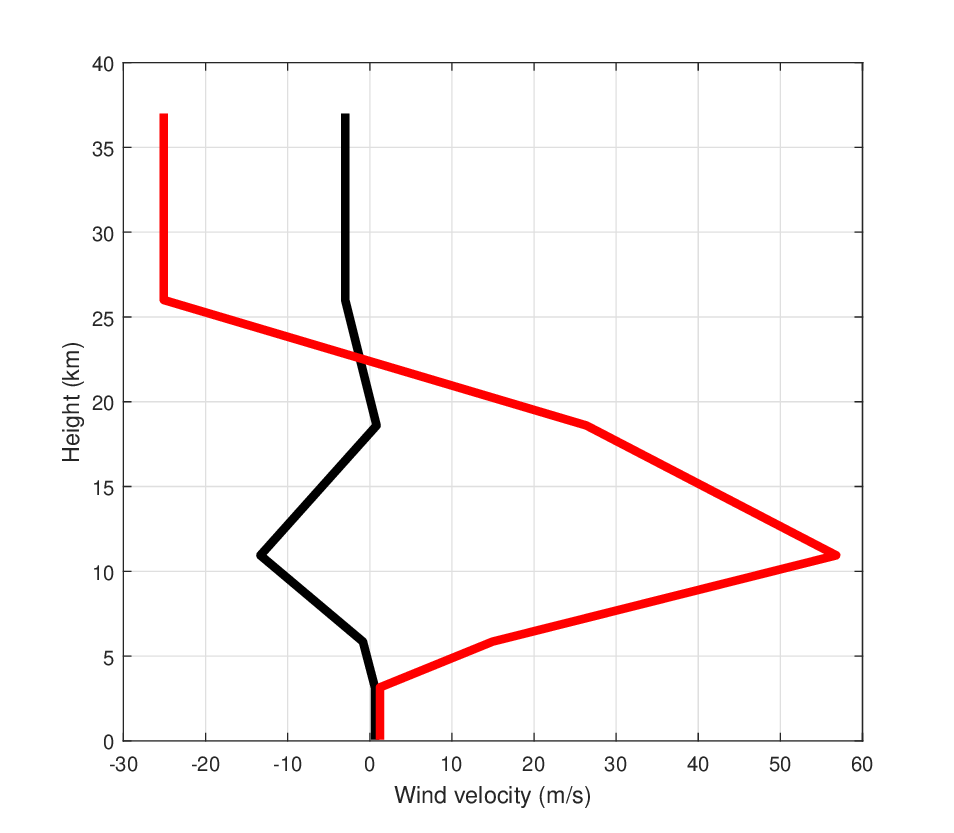}}
\end{center}
\caption{Winds velocity vs. height in the reference meteoroid system from GFS global atmospheric model. Black: fireball motion direction, Red: orthogonal direction. Values greater than 0 are against meteoroid motion}
\label{fig:winds}
\end{figure}

In order to model the dark flight phase it is important to know the atmospheric profile for the time and area closest to the meteoroid fall because the residual meteoroid trajectory, after the end of the luminous path, can be heavily influenced by the atmospheric conditions\cite{Ceplecha1987, Carbognani2020}. We get atmospheric data about pressure, temperature, winds speed and directions by the GFS global atmospheric model\footnote{https://earth.nullschool.net/} near the time and place of the fireball explosion (see Fig.~\ref{fig:winds}). The vertical resolution of this model is height-dependent, from 100 m near the ground to about 7 km around 20 km height, but sufficient to a first raw estimate of the strewn field position with the same method used for the Cavezzo meteorite\cite{Gardiol2020}.\\

\begin{table}[h]
\centering
\caption{Strewn field nominal centers for different mass/section ratio (spherical shape) and a mean meteorite density of about $3500~\textrm{kg}/\textrm{m}^3$. The order of magnitude of the extent of the strewn field for each $M/D$ ratio is also given.}
\label{tab:strewn_field_center}
\begin{tabular}{lccc}
\hline
$M/S$ ($\textrm{kg}/\textrm{m}^2$)      & Lat. N, Long. E ($^\circ$)        & Approx diameter (m) & Strewn field extension (km)\\
\hline
1000                                             & 30.3, 96.5  & 0.4  & $20 \times 100$\\
600	                                             & 30.5, 96.4  & 0.3  & $20 \times 100$\\
300	                                             & 30.8, 96.4  & 0.1  & $20 \times 100$\\
100                                              & 31.3, 96.3  & 0.04 & $20 \times 50$ \\
\hline
\end{tabular}
\end{table}

\begin{figure}[h]
\begin{center}
\resizebox{0.4\textwidth}{!}{\includegraphics{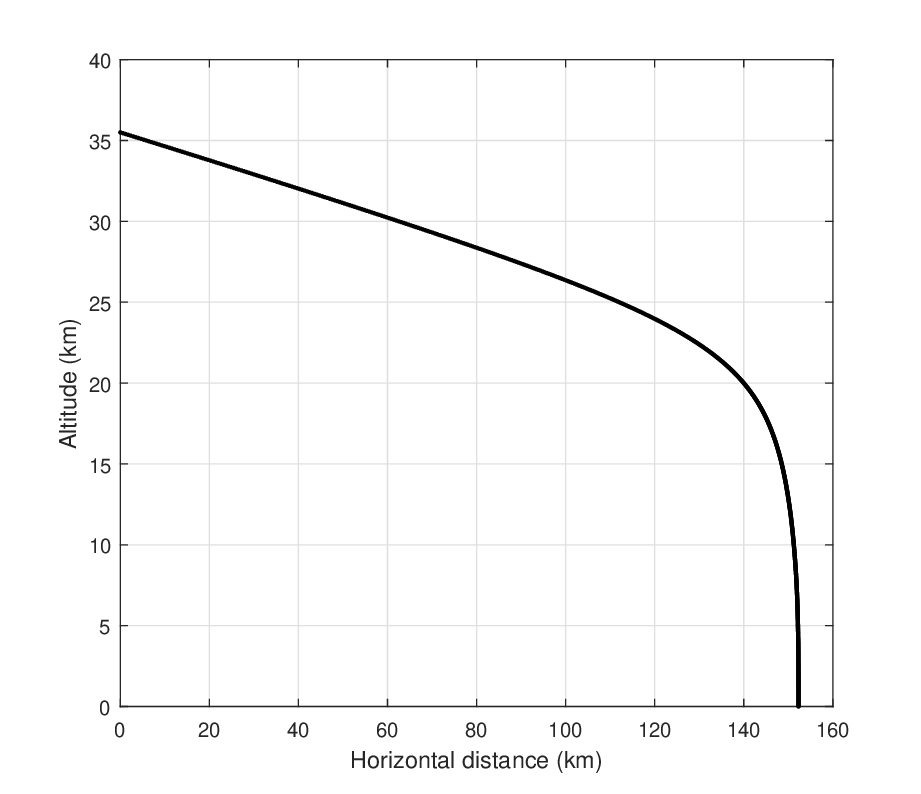}}
\end{center}
\caption{Side view of the nominal dark flight for a $M/S\approx 600 ~\textrm{kg}/\textrm{m}^2$ meteoroid after the main explosion, 35.5 km above the ground. In this simulation the meteoroid continued the fall with the same speed and direction as the original body.}
\label{fig:dark}
\end{figure}

The motion of the residual meteoroids, starting from the observed terminal point of the luminous path, was described using Newton's Resistance law as in Ceplecha \cite{Ceplecha1987}. We assumed that the atmosphere had little effect on the small asteroid, as in Chelyabinsk case\cite{Popova2013}, and that the starting speed of the fragments - after the main explosion - was around the pre-atmospheric value, i.e. $13.6 \pm 1 $ km/s. As a starting point for dark flight computation we have taken the main explosion coordinates given by CNEOS but with different mass/section ratio for the fragments. An average density $\rho \approx 3500~\textrm{kg}/\textrm{m}^3$ - typical of a chondrite - has been assumed for the fragments, because it is reasonable to think that single meteorites have a higher density than the original asteroid.\\
A dark flight example for $M/S\approx 600 ~\textrm{kg}/\textrm{m}^2$ is shown in Fig.~\ref{fig:dark}. In the final part of the dark flight the meteoroid was deflected towards East by wind, which distorted the final trajectory by about 2 km. As the mass/section ratio decreases, the wind influence is greater, so the strewn field location strictly depend from the assumed $M/S$ ratio. The nominal coordinates of the impact points for meteorites with different mass/section ratio are given in Table~\ref{tab:strewn_field_center} and plotted in Fig.~\ref{fig:fireball_trajectory}. Considering the low trajectory inclination and the uncertainties of the explosion point, the strewn field orthogonal extension - with respect to the motion direction - is about $\pm 20$ km. The biggest meteorites, if any, are expected to be in a quite mountainous region, difficult to reach (see Fig.~\ref{fig:fireball_trajectory}). However, given the uncertainties involved and the fact that we don't know how the meteoroid fragmented, any point between $M/S\approx 100~\textrm{kg}/\textrm{m}^2$ and $M/S\approx 1000~\textrm{kg}/\textrm{m}^2$, is eligible for meteorites search. The strewn field total area can be estimated in about 4000 $\textrm{km}^2$. 
Obviously this area is reduced if we look for meteorites with a well defined $M/S$ ratio (see Table~\ref{tab:strewn_field_center}).

\section{Conclusions}
\label{sec:end} 
We give some results about the CN20201222 fireball, observed at about 23h 23m 33s UTC on December 22, 2020 and detected by the US surveillance satellites. The fireball was generated by the fall of a small asteroid of about 7 m in diameter. The explosion height is consistent with a fragmented rocky body and the computed heliocentric orbit supports an asteroidal origin. Fortunately, the potential fragments of the small asteroid did not fall in sea so, with an accurate search on the ground, it will be possible to find some interesting samples of the original body. The probable total strewn field is approximately between two points with geographical coordinates ($+31.3^\circ$ N; $96.3^\circ$ E) and ($+30.3^\circ$ N; $96.5^\circ$ E), for a surface area of about 4000 $\textrm{km}^2$. This large extension is a direct consequence of the low inclination of the trajectory, fortunately the strewn field for selected $M/S$ ratio are much smaller.\\
Given the unknown uncertainty of the satellites data - we have assumed reasonable uncertainties values based on some well-studied past cases - these results should be taken with caution.
In this regard, it would be very appropriate to add the estimate of the uncertainties to the satellite data released by the CNEOS which, however, are of military origin. Observations from orbit of atmospheric explosions would be a powerful tool for studying the origin of bright fireballs and could be very useful for delimiting strewn fields on the ground. Based on our experience with CN20201222, if the data are also to be useful for the meteorites search, the uncertainties should be about an order of magnitude lower than the values adopted here.

\section*{Acknowledgements}
Many thanks to Fabrizio Bernardi (SpaceDys), for the fast computing of the NEAs encounter conditions with Earth used in Section~\ref{sec:orbit_progenitor}.
%
%

\end{document}